\documentclass[conference]{IEEEtran}
\IEEEoverridecommandlockouts

\usepackage{cite}
\usepackage{amsmath,amssymb,amsfonts}
\usepackage{graphicx}
\usepackage{textcomp}
\usepackage{xcolor}
\usepackage{multirow}
\usepackage{algorithm}
\usepackage{algorithmic}
\usepackage{booktabs}
\usepackage{url}
\usepackage{array}
\usepackage{soul}

\usepackage[letterpaper, top=0.75in, bottom=1.05in, left=0.65in, right=0.65in]{geometry}

\def\BibTeX{{\rm B\kern-.05em{\sc i\kern-.025em b}\kern-.08em
    T\kern-.1667em\lower.7ex\hbox{E}\kern-.125emX}}
\title{Jamming Smarter, Not Harder: Exploiting O-RAN Y1 RAN Analytics for Efficient Interference}
\author{
Abiodun Ganiyu\textsuperscript{1},
Dara Ron\textsuperscript{1},
Syed Rafiul Hussain\textsuperscript{2}, and
Vijay K Shah\textsuperscript{1}\\
\textsuperscript{1}NextG Wireless Lab, North Carolina State University, Raleigh, USA \\
\textsuperscript{2}Computer Science and Engineering, The Pennsylvania State University, Pennsylvania, USA \\
\{aganiyu, dron, vijay.shah\}@ncsu.edu, hussain1@psu.edu
}

\begin{document}

\maketitle

\begin{abstract}
The Y1 interface in O-RAN enables the sharing of RAN Analytics Information (RAI) between the near-RT RIC and authorized Y1 consumers, which may be internal applications within the operator's trusted domain or external systems accessing data through a secure exposure function. While this visibility enhances network optimization and enables advanced services, it also introduces a potential security risk -- a malicious or compromised Y1 consumer could misuse analytics to facilitate targeted interference. In this work, we demonstrate how an adversary can exploit the Y1 interface to launch selective jamming attacks by passively monitoring downlink metrics. We propose and evaluate two Y1-aided jamming strategies: a clustering-based jammer leveraging DBSCAN for traffic profiling and a threshold-based jammer. These are compared against two baselines strategies -- always-on jammer and random jammer -- on an over-the-air LTE/5G O-RAN testbed. Experimental results show that in unconstrained jamming budget scenarios, the threshold-based jammer can closely replicate the disruption caused by always-on jamming while reducing transmission time by 27\%. Under constrained jamming budgets, the clustering-based jammer proves most effective, causing up to an 18.1\% bitrate drop while remaining active only 25\% of the time. These findings reveal a critical trade-off between jamming stealthiness and efficiency, and illustrate how exposure of RAN analytics via the Y1 interface can enable highly targeted, low-overhead attacks, raising important security considerations for both civilian and mission-critical O-RAN deployments.

%Results show that under an unconstrained setting, the threshold-based jammer can closely mirror the impact of always-on jamming while reducing transmission time by 27\%. Under constrained jamming budgets, the clustering-based jammer achieves the highest disruption when targeting high-throughput traffic, demonstrating up to 18.1\% bitrate drop with only 25\% jamming activity. These findings reveal a key trade-off between jamming aggressiveness and efficiency, and highlight how exposed RAN analytics via Y1 interface can enable precision attacks with minimal transmission effort raising important considerations for O-RAN security in both civilian and mission-critical deployments. 

\end{abstract}

\begin{IEEEkeywords}
Stealthy Jamming attacks, RAN Analytics Information, Y1, O-RAN, 5G
\end{IEEEkeywords}

\section{Introduction}
The Open Radio Access Network (O-RAN) paradigm is reshaping the telecommunications industry by promoting openness, modularity, and vendor interoperability. Through standardized interfaces and disaggregated components, O-RAN enables flexible integration of third-party applications and intelligent control mechanisms across the RAN \cite{tripathi2025fundamentals}.
%\cite{oran-ml2, oran1}. 
This architectural shift supports AI-driven network automation, fine-grained optimization, and cost-effective network customization, capabilities critical to commercial networks and tactical and defense communications systems. The architecture’s support for low-latency control, scalable radio coordination, and dynamic analytics enables future battlefield networks to operate with enhanced situational awareness, rapid network reconfiguration, and resilient command-and-control.

The relevance of O-RAN to military communications is gaining traction. Defense agencies and researchers are increasingly exploring how O-RAN’s software-defined architecture can enable resilient, secure, and mission-adaptive networks \cite{dualuse2024}. The Department of Defense (DoD) and other allied institutions have identified the potential for dual-use platforms, where commercial-grade O-RAN systems can be adapted to meet military-grade requirements such as hardened security, real-time responsiveness, and contested spectrum operation \cite{nist8560}. In particular, O-RAN’s support for intelligent RAN analytics, distributed control, and programmable policy enforcement aligns well with operational needs on future battlefields.

The O-RAN ALLIANCE continues to evolve this vision by releasing specifications for key interfaces such as the O-fronthaul, F1, E2, A1, and O1, which standardize connectivity across RAN components and intelligent controllers \cite{oran_specifications}. One notable development is the Y1 interface, introduced by the O-RAN ALLIANCE Working Group 3~\cite{oran_y1}. The Y1 interface enables the exposure of near-real-time RAN Analytics Information (RAI) from a Y1 Producer (co-located with the near-RT RIC) to authorized Y1 consumers which may be internal applications within the operator’s trusted domain or external systems accessing data through a secure exposure function. This interface is envisioned to enable a wide range of third-party analytics services, such as traffic orchestration, V2X coordination, and mission-critical awareness platforms, by delivering near-real-time metrics from the RAN. These capabilities are especially relevant in defense and emergency communications, where accurate and timely RAN insights may drive adaptive spectrum use, network resilience, and coordinated mobility. However, this increased visibility also introduces new attack vectors. A compromised or malicious Y1 consumer could misuse exposed analytics to infer network behavior and strategically disrupt critical operations.

Motivated by this threat, our work investigates the feasibility of RAN analytics-driven efficient interference through the Y1 interface. We demonstrate how an adversary, operating as an authenticated but malicious Y1 consumer, can access RAN metrics streamed over Y1 to infer traffic patterns and coordinate targeted jamming attacks. Specifically, the Y1 consumer forwards real-time analytics to an external software-defined (or GNU Radio) based jammer, which then selectively activates interference based on observed traffic conditions. By leveraging unsupervised learning (via clustering technique) to profile traffic, the jammer intelligently allocates limited transmission resources to maximize disruption, a tactic particularly relevant for covert or energy-constrained military operations.

While prior efforts have explored various machine learning (ML) strategies for O-RAN security enhancement \cite{balakri, 10416344, 10971998, 10697477}, this work highlights a lesser-discussed threat: \textit{the use of exposed RAN analytics as a side channel for targeted radio interference}. We implement our attack on an over-the-air testbed and compare three strategies, always-on, random, and traffic-aware jamming, demonstrating how analytics-guided attacks can achieve comparable disruption with significantly less transmission effort. Our findings underscore the urgent need to consider threat models involving analytics exposure, especially in sensitive deployments such as defense and tactical communications, where system compromise or data leakage may have high operational costs. 

The key contributions of this work are outlined as follows.

$\bullet$ We demonstrate how a threat actor can exploit the O-RAN Y1 interface to passively monitor RAN Analytics Information (RAI) and gain situational awareness of network activity. Using this visibility, we illustrate the feasibility of selective jamming attacks driven by unsupervised traffic profiling using DBSCAN, enabling context-aware interference without requiring privileged access to the RAN itself.

$\bullet$ We develop a complete O-RAN testbed integrating a compliant Y1 interface, srsRAN-based RAN/Core systems, and GNU Radio based jamming implementations. Our threat model emulates a malicious Y1 consumer relaying RAI to a jammer, which dynamically adjusts its activity based on learned traffic patterns and jamming constraints.

$\bullet$ We conduct two complementary experiments: one under unlimited jamming budget using a fixed traffic pattern, and another under constrained jamming budgets with multi-rate traffic. Results show that threshold-based and DBSCAN-guided jammers can closely mirror the disruptive effect of always-on jamming with significantly less transmission time, and that targeting high-throughput traffic yields the greatest impact per unit of jamming effort.

\section{Related Works}
As O-RAN adoption grows, the security implications of its open and modular design have drawn increasing attention from both academia and industry. A prominent concern is the expanded attack surface introduced by programmable RAN controllers and third-party applications (xApps), particularly within the Near-real-time RIC environment.

Several prior works have focused on enhancing O-RAN security through machine learning (ML)-based xApps. For example, \cite{balakri} presents an ML-driven connection management xApp capable of both launching and defending against malicious behavior. A federated deep reinforcement learning (FDRL) framework for detecting jamming attacks is proposed in \cite{10416344}, while other works explore DRL-based resource allocation strategies to counter adversarial interference \cite{10971998,10697477}. Additional approaches include classifier xApps that identify and mitigate jamming by classifying interference types \cite{chiejina2024}, intrusion detection mechanisms for threat identification and mitigation \cite{moore2025}, and anomaly detection systems that protect the near-RT RIC from attacks targeting the E2 interface \cite{10907911}. These studies primarily address adversarial behavior on internal O-RAN interfaces (e.g., E2, A1, and F1), focusing on attacks and defenses involving internal control loops or xApps. To the best of our knowledge, no existing work has explored attack scenarios involving an untrusted or malicious Y1 consumer, a new telemetry exposure point in the O-RAN architecture that grants visibility into real-time RAN analytics.

Beyond direct attack mitigation, recent efforts have examined the security of third-party components in O-RAN systems. For example, Atalay et al. \cite{atalay2023} introduce the xApp Repository Function (XRF) to address authentication and authorization challenges when integrating external xApps. Similarly, the work in \cite{aizikovich2025roguecelladversarialattack} explores adversarial activity via the Rogue Cell attack, where malicious operators manipulate RAN telemetry to mislead traffic steering decisions. Their proposed attack (APATE) demonstrates how such manipulation can lead to unfair resource allocation, while their LSTM-based detection mechanism (MARRS) offers a potential countermeasure. In parallel, other studies have explored jamming strategies in wireless networks. One such effort presents a systematic learning method for optimal jamming, using reinforcement learning to adapt jamming behavior based on channel conditions~\cite{amuru2025}.

In contrast to these efforts, our work is the first to investigate how a compliant yet malicious Y1 consumer can be exploited to coordinate external analytics-guided jamming. We demonstrate how passive access to RAN Analytics Information (RAI) without breaching the RAN or modifying internal xApps can enable an attacker to infer traffic behavior and drive targeted interference via a coordinated external jammer. This expands the O-RAN threat model to account for telemetry leakage and adversarial use of exposed interfaces, particularly under the evolving Y1 specification.

\section{O-RAN Background}
We briefly outline the key components of the O-RAN architecture relevant to our work. Readers should refer to \cite{tripathi2025fundamentals} for a detailed understanding of O-RAN architecture.

\subsection{RAN Intelligent Controller (RIC)}
The RAN Intelligent Controller (RIC) is a central component introduced in O-RAN to support programmability and intelligent control of the radio access network. It comprises two logical components based on operational time scales: the near-real-time RIC and the non-real-time RIC.

\noindent $\bullet$ \textbf{Near-Real-Time RIC:} Operates within 10 ms to 1 s timescales. It hosts xApps that perform latency-sensitive control and optimization tasks, such as handover management or interference mitigation. It communicates with the underlying RAN nodes via the E2 interface and supports telemetry gathering and control actuation.

\noindent $\bullet$ \textbf{Non-Real-Time RIC:} Operates on a time scale of greater than 1 s and up to minutes, handling long-term network optimization tasks, such as policy generation and model training, hosted within the Service Management and Orchestration (SMO) framework. It connects to the near-RT RIC via the A1 interface and to other O-RAN components, such as the O-DU, through the O1 interface.

\subsection{RIC Database and Shared Data Layer}
The RIC database serves as a centralized database repository that stores various forms of RAN state information, including Key Performance Measurements (KPMs), RAN events, and custom cell-level and UE-level analytics. xApps running on the near-real-time RIC can read from and write to this database to support data-driven decision-making. 

Access to the RIC database is mediated through the Shared Data Layer (SDL), which provides a standardized API for storing and retrieving analytics data. The SDL, as implemented by the O-RAN Software Community (OSC) \cite{osc-sdl}, interfaces with a Redis backend and exposes Redis-compatible functions through a controlled API surface. This abstraction allows xApps to interact with the database using familiar key-value operations without direct access to the database internals. 

Importantly, the Y1 Producer also interacts with the SDL to access stored analytics data. When a Y1 Consumer subscribes to specific RAN metrics or analytics, the producer queries the SDL and constructs a response conforming to the Y1 specification, ensuring only authorized and subscribed consumers receive the requested analytics.

\subsection{Y1 Interface}
%The Y1 interface is a newly standardized interface introduced by the O-RAN Alliance to facilitate the delivery of RAI from the Near-RT RIC to authorized external entities known as Y1 Consumers. Its primary objective is to enable data-driven network intelligence and situational awareness for advanced use cases such as vehicular communications (V2X), real-time diagnostics, and cross-layer optimizations.  

%(Dara: This sentence was already introduced in the Introduction section. We should delete it to avoid duplication and reduce the page length.)

%At a high level, 
The Y1 interface supports two modes of RAI access: (1) subscription-based notification, where the Y1 Consumer registers interest in specific analytics and receives asynchronous updates; and (2) query-based retrieval, where the Y1 Consumer explicitly requests specific analytics on demand. These operations are defined through three core APIs: 1) \textit{RAI\_Subscribe} initiates a subscription to a particular RAI type and target; 2) \textit{RAI\_Notify} delivers the analytics payload based on the notification method (periodic or event-driven); and 3) \textit{RAI\_Query} responds with RAI data upon request.

The analytics content, as defined under the ``RAN performance analytics" type in the Y1 specification, includes a set of parameters such as average RLC throughput, downlink latency, packet loss rate, and their distributions. Importantly, while the specification allows analytics to be per-UE or slice-specific, in this study, we scope our use of RAN analytics to aggregated cell-level metrics to avoid user privacy concerns and ensure generality. 

Internally, the Y1 Producer, which resides within the Near-RT RIC, sources these analytics from the RIC Database via the SDL. It then filters and encodes the data based on the subscription query or request filter, returning only the subset of attributes as requested by the Y1 Consumer. 
The protocol stack for Y1 follows REST over HTTPS (as per the O-RAN specification \cite{oran_y1}), utilizing JSON for data interchange. Mandatory security controls, including mutual TLS (mTLS) for authentication and authorization, are enforced to restrict access only to verified and permitted consumers. However, the specification does not define mechanisms for fine-grained auditing or behavioral analysis of consumers post-authentication, leaving room for potential misuse.

This paper focuses on the RAI\_Subscription operation with periodic notification trigger, which allows the attacker-controlled consumer to receive a continuous stream of analytics updates at configured intervals (e.g., every 1s or 5s). The flexibility of this subscription mechanism, while beneficial for analytics-driven services, introduces a stealthy reconnaissance channel if abused by malicious or compromised entities.

\begin{figure}[t]
    \centering
    % \vspace{-0.25in}
    % \includegraphics[width=1.15\linewidth]{images/y1-illus.pdf}
    \includegraphics[width=1\linewidth, trim={1cm 0.8cm 4cm 1cm}, clip]{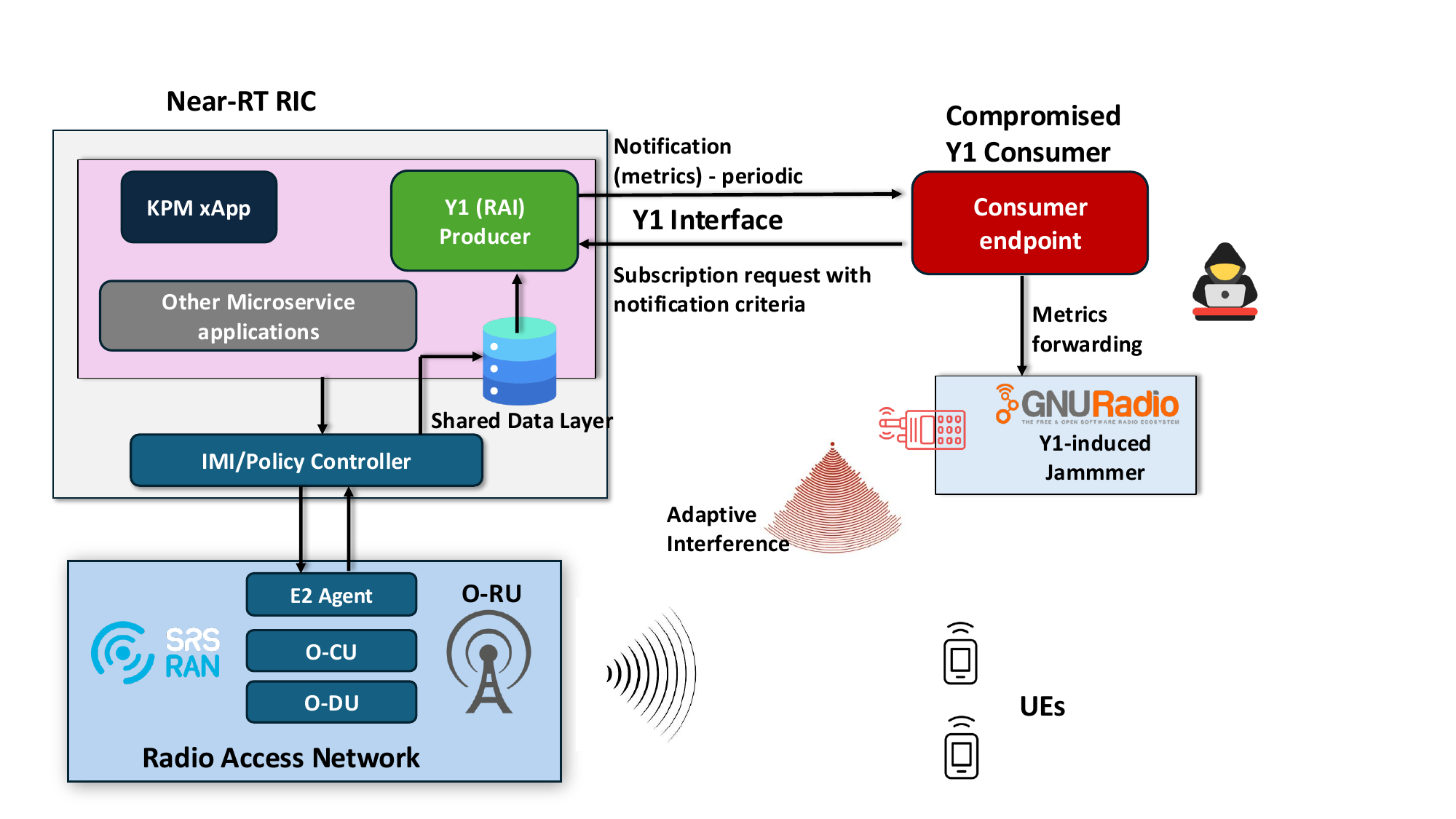}
    \vspace{-0.15in}
    \caption{Threat model. A malicious Y1 Consumer leverages legitimate access to RAN analytics to coordinate adaptive jamming against the RAN.}
    \label{fig:system}
    \vspace{-0.2in}
\end{figure}

\section{Threat and System Model}
\subsection{Threat Model}

As shown in Fig. \ref{fig:system}, the RAN exposes telemetry data (e.g., KPMs) to the near-RT RIC via the E2 interface. These metrics are processed by internal xApps (e.g., KPM xApp) and stored within a central RIC Database. The RAI Producer, implemented in compliance with the Y1 interface specification, exposes these analytics to authorized Y1 consumers based on defined subscription criteria. Notifications may be event-triggered or periodic, and are securely exchanged using mutual TLS (mTLS) authentication.
%In this work, we consider an external threat scenario involving a malicious actor operating as an authorized Y1 Consumer. While this entity resides outside the near-real-time RIC, it has been legitimately authenticated and subscribed via the Y1 interface, as specified in the O-RAN Y1 specification. 
The attacker does not attempt to compromise the RIC or RAN infrastructure directly but instead leverages their authorized access to RAN analytics to facilitate external interference operations. Specifically, the Y1 Consumer, while appearing benign and authenticated, is compromised or operated by a malicious insider who forwards the received RAN analytics to a coordinated external jammer. This leads to a novel form of \textit{analytics-driven jamming}, where the interference is intelligently triggered based on accurately observed near-real-time RAN performance data.

The attack model assumes: \textit{1. The Y1 Consumer has valid credentials and is able to establish a mutual TLS connection with the Y1 Producer; 2. The analytics shared are compliant with the consumer’s subscription and the producer’s export policies; 3. The jammer is physically decoupled from the RIC but receives continuous analytics from the compromised consumer.} This threat highlights the risks associated with exposing aggregated RAN telemetry to third-party consumers, even when such access is governed by standard authentication mechanisms. It underscores a critical tension between openness for innovation and the potential for misuse of authorized visibility within the O-RAN architecture.

%\subsection{System Overview}
%Our system architecture models an O-RAN deployment where analytics are streamed from the RAN to external consumers via the standardized Y1 interface. The setup involves five main entities: the Radio Access Network (RAN), the near-RT RIC, a legitimate but compromised Y1 consumer, a malicious controller, and a jammer.

In our threat scenario, the Y1 consumer is compromised. While still operating within its access limits (for example, receiving only authorized RAN analytics), it forwards the metrics received to an external malicious agent, a controller responsible for managing a physical jammer. The jammer uses this analytics stream to dynamically adapt its interference behavior to degrade network performance in a stealthy and energy-efficient manner.

\begin{table}[t]
\tiny
\centering
\caption{Implemented Y1 Interface API Summary}
\begin{tabular}{|l|l|p{4cm}|}
\hline
\textbf{Endpoint} & \textbf{Method} & \textbf{Description} \\
\hline
\texttt{/subscriptions/subscribe} & POST & Registers a Consumer to receive RAN analytics \\
\texttt{/subscriptions/unsubscribe} & DELETE & Cancels an existing subscription \\
\texttt{/subscriptions/\textless id\textgreater} & PUT & Updates an existing subscription (partial support) \\
\texttt{/notify} (Consumer-side) & POST & Receives metrics from the Producer \\
\hline
\end{tabular}
\vspace{-0.2in}
\end{table}

\subsection{Y1 Interface Implementation}
To emulate a real-world compliant environment, we implemented the Y1 interface as defined in the O-RAN ALLIANCE Y1 specifications (Y1TD, Y1AP, Y1GAP). Our design includes both the Y1 Producer (within the Near-RT RIC) and the Y1 Consumer, ensuring mutual Transport Layer Security (mTLS)-based authentication and structured data exchange~\cite{oran2024security}. Since the current open-source near-RT RIC platforms, including the O-RAN Software Community (OSC) implementation, do not yet support the Y1 interface, we developed a custom lightweight Y1 Producer using Flask in Python. This component retrieves analytics from a Redis-backed SDL and exposes the RAI to authenticated consumers over a secure mTLS channel. To enable integration with the near-RT RIC platform, we containerized the Y1 Producer as a Docker image which can be deployed as a Kubernetes service within the RIC environment. This mirrors real-world deployment practices and ensures seamless communication with other RIC platform components. The Y1 Consumer, also implemented as a Flask web service, served as the interface through which RAI was forwarded to an external jammer in our threat model.

\subsubsection{Producer API Design}
The Producer exposes a secure RESTful interface, including a subscription endpoint at \texttt{\small /Y1\_RAI\_Subscriptions/v1/subscriptions/subscribe}, through which authenticated Consumers can submit subscription requests. Each request includes:

\noindent $\bullet$ \texttt{\small raiType} and \texttt{\small raiTypeVersion}: defining the category of analytics,

\noindent $\bullet$ \texttt{\small notificationCriteria}: specifying the trigger mechanism (periodic or event-based) and the interval,

\noindent $\bullet$ \texttt{\small notificationTargetAddress}: indicating the Consumer endpoint for metric delivery.

Upon subscription, the Producer retrieves analytics from the Redis-backed SDL and transmits them to the Consumer based on the subscription parameters. The metrics are JSON-encoded and include; \texttt{\small subscription\_id}, \texttt{\small rai\_content}, \texttt{\small timestamp}, and \texttt{\small validity\_period}.

\subsubsection{Consumer API Implementation}
The Y1 Consumer is implemented as a secure Flask-based HTTPS service that initiates a subscription to the Y1 Producer using mTLS with X.509 certificate-based authentication, in compliance with O-RAN security guidelines. Upon successful subscription, the Consumer exposes a notification endpoint that periodically receives the RAI pushed by the Producer. All incoming RAI messages are parsed and stored in memory for subsequent use. To emulate an adversarial scenario, the Consumer includes a malicious forwarding module which establishes a TCP socket with the jammer system and continuously streams the latest received RAI metrics.

We leverage the KPM xApp within the Near-RT RIC to collect performance metrics from the RAN over the E2 interface. These metrics are aggregated periodically and stored in a Redis database, where they are accessed by the Y1 Producer to subsequently serve subscribed Y1 consumers. Table~\ref{tab:analytics-metrics} summarizes the nine metrics extracted and their relevance to RAN behavior and performance.

\begin{table}[t]
\tiny
\centering
\caption{Extracted RAN Analytics Metrics for Y1 Interface}
\label{tab:analytics-metrics}
\begin{tabular}{|l|p{6cm}|}
\hline
\textbf{Metric} & \textbf{Description} \\
\hline
DL CQI & Average Channel Quality Indicator from UEs, indicating link quality. \\
\hline
DL MCS & Average Modulation and Coding Scheme used for downlink scheduling. \\
\hline
DL Bitrate & Total downlink throughput (in bits/sec) across all active UEs. \\
\hline
DL BLER & Downlink Block Error Rate, computed from transmission error statistics. \\
\hline
DL Latency & Average latency (in seconds) from the PDCP layer. \\
\hline
DL Bytes & Total number of acknowledged downlink bytes from PDCP. \\
\hline
PCI & Physical Cell Identifier of the transmitting cell. \\
\hline
Carrier ID & Logical carrier index, useful in multi-carrier environments. \\
\hline
Number of RACH & Count of random access procedure attempts (Msg1 preambles). \\
\hline
\end{tabular}
\vspace{-0.2in}
\end{table}

\section{Y1-Aided Jamming Strategies}

\subsection{Y1-Threshold Jammer}
The Y1-Threshold jammer is a lightweight yet effective strategy that reacts to real-time RAN metrics streamed via the Y1 interface. 
%Unlike clustering-based models, it does not require offline training. Instead, 
Specifically, it applies a simple thresholding rule on selected features (e.g., CQI, bitrate, or BLER) to decide when to transmit. This design is particularly useful for constrained systems where compute or memory overhead must be minimized.
% Jamming Decision Logic 
\noindent Let $\mathbf{x}_t = \{\text{CQI}_t, \text{MCS}_t, \text{Bitrate}_t, \text{BLER}_t\}$ represent the observed downlink analytics vector at time $t$. The jammer compares selected features against a threshold:
\begin{equation}
\mathrm{JAM} \quad \text{if } \text{Bitrate}_t \geq \theta; \quad \mathrm{NO\ JAM} \quad \text{otherwise},
\end{equation}
where $\theta$ is a predefined threshold selected based on offline traffic profiling. For example, if the traffic is expected to operate consistently at $4$ Mbps during active periods, a threshold $\theta = 1$ Kbps may be sufficient to distinguish active from idle states. This threshold-based mechanism enables the jammer to concentrate energy on periods of legitimate traffic, reducing overall transmission time and enhancing stealth. 
% As demonstrated in our fixed traffic scenario (Part A), the threshold jammer closely approximates the disruption caused by an always-on jammer while reducing jamming activity by over 25\%.

\subsection{Clustering-based Jammer}
% \underline{DBSCAN Algorithm } %\Dara{ Need the Discussions}} 
%We design and implement a clustering-based jamming strategy where the jammer passively monitors RAN analytics streamed by a subscribed and authenticated Y1 Consumer. Over a period of observation, it collects traffic feature vectors and uses them to train a clustering model offline. The trained model is then leveraged for live classification: at runtime, the jammer classifies analytics samples and selectively activates interference based on the inferred traffic class. \textit{This enables a stealthy and adaptive jamming strategy that targets only meaningful traffic patterns, conserving energy and avoiding detection.}

We design and implement a clustering-based jamming strategy in which the jammer passively monitors RAN analytics streamed by a subscribed and authenticated Y1 consumer. Specifically, during an observation phase, it collects traffic feature vectors and trains a clustering model offline. This model is then used for real-time classification: at run-time, the jammer analyzes incoming analytics samples, infers their traffic class, and selectively activates interference accordingly. This approach enables a stealthy and adaptive jamming strategy that targets only high-value traffic patterns, improving jamming efficiency and minimizing the risk of detection.

\textit{Offline Training (Clustering Phase)} in DBSCAN is applied to determine the optimal centroids of valid clusters, denoted by $c_j^*$, which are then used during \textit{Online Testing} to identify whether the network is jammed. Let $\mathcal{X} = \{\mathbf{x}_1, \mathbf{x}_2, \dots, \mathbf{x}_n\}$ be the set of KPMs collected from a real testbed, where $\mathbf{x}_i \in \mathbb{R}^d$ denotes the $i$-th traffic feature vector. In our case, each vector $\mathbf{x}_i$ is defined as a set $\{\text{CQI}, \text{MCS}, \text{Bitrate}, \text{BLER}\}$.
%\paragraph{Offline Training (Clustering Phase)}
%Let $\mathcal{X} = \{\mathbf{x}_1, \mathbf{x}_2, \dots, \mathbf{x}_n\}$, where $\mathbf{x}_i \in \mathbb{R}^d$ denotes the $i$-th traffic feature vector. In our case, eqach vector is defined as $\mathbf{x}_i = [\text{CQI}, \text{MCS}, \text{Bitrate}, \text{BLER}]$.
The standard score normalization is given by
$\hat{\mathbf{x}}_i = (\mathbf{x}_i - \mu ) / \sigma, \quad \forall i \in \{1, \dots, n\},$
where $\mu$ and $\sigma$ are the mean and standard deviation vectors computed per feature. 

We apply the unsupervised learning algorithm DBSCAN, as described in \cite{PhysRevE.110.045207}, to cluster the feature vectors $x_i$, by requiring that the n-th nearest neighbor of a point lies within a specified distance $\varepsilon$. Points for which this distance exceeds $\varepsilon$ are considered noise or outliers, and do not belong to any cluster. The DBSCAN algorithm can be expressed as follows:
\begin{align}
\texttt{DBSCAN}(\hat{\mathcal{X}}, \varepsilon, \texttt{minPts}) & \nonumber \\
 \xrightarrow{\text{Output}} \mathcal{L} =  \{l_1, \dots , & l_n\},  \quad l_i \in \{-1, 0, \dots, k-1\},
\end{align}
where $\hat{\mathcal{X}}$ denotes the normalized KPMs, and $\texttt{minPts}$ is the minimum number of points required to form a dense region. Each $l_i$ assigns a point to a cluster, and $l_i = -1$ indicates noise. The optimal centroids of valid clusters (excluding noise) are computed as: $\mathbf{c}_j^* = \frac{1}{|\mathcal{C}_j|} \sum_{\hat{\mathbf{x}}_i \in \mathcal{C}_j} \hat{\mathbf{x}}_i, \quad \forall j \in \{0, \dots, k-1\}$, where $|\mathcal{C}_j| \ge \texttt{minPts}$ is the number of points assigned to cluster $j$.

%\textit{Online Inference (Live Classification):} 
\textit{Online Testing (Live Classification):}  At runtime, the jammer receives a new vector $\mathbf{x}_t$ and performs the following procedures to identify jamming traffic. First, the feature vector is normalized using the $\texttt{scaler.transform}(\cdot)$ function, resulting in $\hat{\mathbf{x}}_{t}$. Next, the optimal centroids $c_j^*$, obtained from the \textit{Offline Training}, are used to assign $\hat{\mathbf{x}}_{t}$ to the nearest cluster $\mathcal{I}$ according to: $j^* = \arg\min_{j} \|\hat{\mathbf{x}}_t - \mathbf{c}_j\|_2$. A jamming decision is then made based on the cluster index:
\begin{align}
    \mathrm{JAM} \quad \text{if } j^* \notin \mathcal{I}; \quad \mathrm{NO\ JAM} \quad \text{if } j^* \in \mathcal{I},
\end{align}
where $\mathcal{I}$ is the index set of clusters designated for jamming based on the strategy (e.g., low or high traffic targeting).

\section{Testbed Development and Performance Evaluation}

\subsection{Experimental Testbed}\label{sec:testbed}

\begin{figure}[!t]
\centering
    \includegraphics[width=0.8\linewidth, trim={5cm 1.5cm 5cm 5cm}, clip]{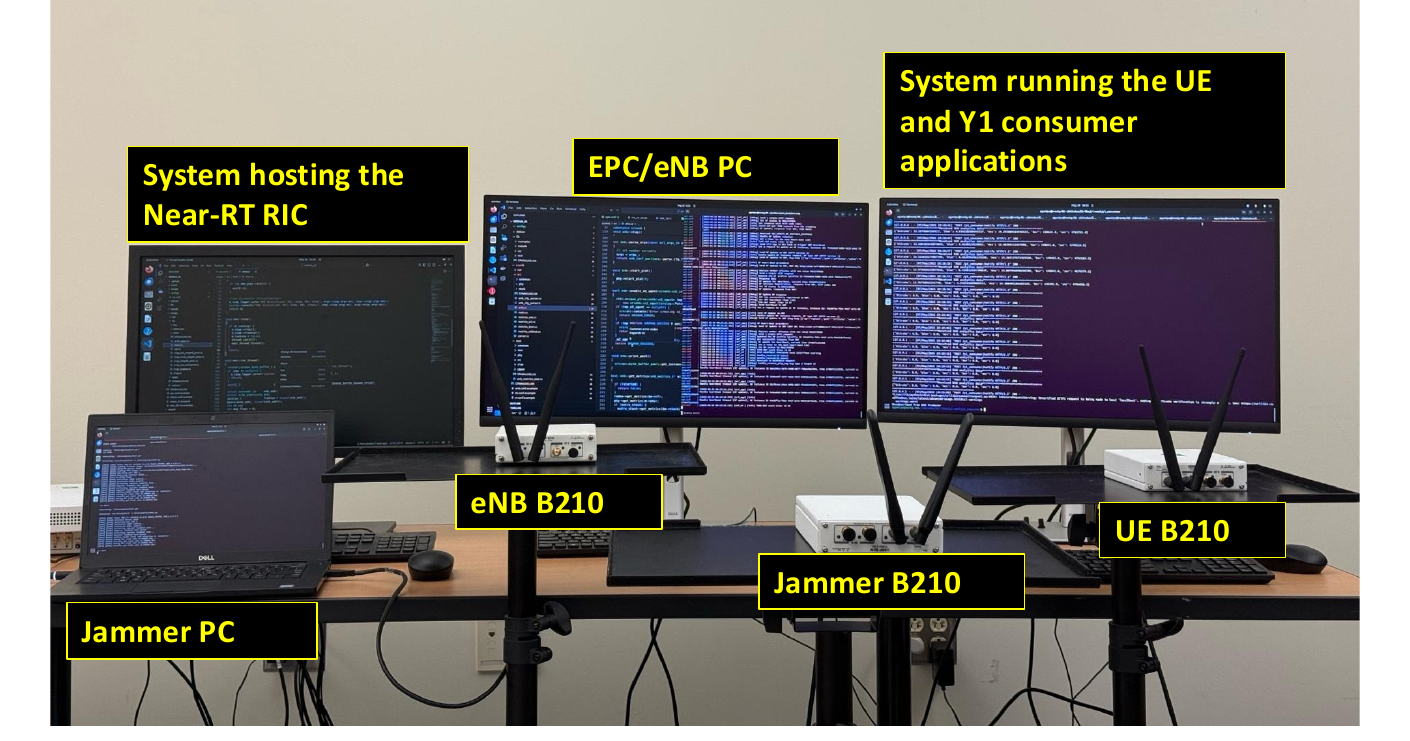}
\caption{This is the OTA experimental setup, which consists of Near-RT RIC, BS, UE, and the jammer using USRP B210s as RF frontends.}
\label{fig:testbed}
\vspace{-0.22in}
\end{figure}

Our experimental testbed, depicted in Fig. \ref{fig:testbed}, comprises four physical systems configured to emulate a realistic O-RAN environment. These systems represent: (i) the RAN and Core Network (BS/EPC), (ii) User Equipment (UE), (iii) Near-RT RIC, and (iv) an external jammer. Three of the four systems are equipped with USRP B210 SDRs for over-the-air (OTA) transmission and reception.

The RAN is based on the srsRAN 4G stack (v21.10) \cite{srsRAN}, with the eNodeB and EPC co-located on a high-performance workstation (Intel Core i9-14900K @ 5.7GHz, NVIDIA RTX 4090 GPU, 64 GB RAM, Ubuntu 22.04). The UE is hosted on a separate system with identical hardware and software configuration, also equipped with a USRP B210 radio frontend. The malicious Y1 Consumer, implemented as a Flask-based API application, is co-located on the UE system to receive analytics from the Y1 producer and relay it to the jammer.

The Near-RT RIC based on Open Software Community (OSC) RIC runs on a dedicated Alienware Aurora R16 (Intel i9-14900KF @ 5.9GHz, NVIDIA RTX 4090, 64 GB RAM, Ubuntu 22.04) which host the Y1 RAI Producer application and a custom-developed KPM xApp that collects downlink RAN performance metrics via the E2 interface. These metrics are published to a Redis-backed SDL, from which the Y1 Producer exposes RAN analytics to authenticated third-party consumers via mTLS-secured API endpoints, adhering to the O-RAN WG3 Y1 specification. In our setup, the notification period which  defines how often the Y1 Producer sends analytics updates to consumers was fixed at 1 second.

To emulate the threat model, we designate a separate laptop (Dell Latitude 7490, Intel Core i7-8650U @ 4.2GHz, 8 GB RAM, Ubuntu 22.04) as the external jammer platform. The jammer is implemented using GNU Radio and supports multiple jamming strategies, including: \textbf{Y1-aided Jamming} -- \textit{Threshold and clustering-based jamming techniques} presented in Section V, and two baseline jamming techniques: \textbf{Always-on Jamming} -- the Jammer transmits continuously across the entire experiment, regardless of traffic activity, and \textbf{Random Jamming} -- randomly activates the jammer in discrete bursts over time, without awareness of network conditions.

We used conventional DSP blocks, including a signal source, throttle, and a USRP sink block. Jamming control is achieved by programmatically adjusting the amplitude of the waveform (set to 1.0 for active jamming and 0.0 otherwise), and dynamically toggling the USRP antenna gain between 31 and 0. While amplitude controls waveform power, the antenna gain setting further amplifies or suppresses transmission. These controls are driven by the jamming logic (threshold or clustering-based) in real time, enabling the jammer to activate or deactivate based on incoming analytics data.

The threshold condition and DBSCAN algorithm are implemented directly within the GNU Radio script, alongside a TCP socket-based module that enables live streaming of RAN analytics from the malicious Y1 Consumer to the jammer. Depending on the selected jamming strategy, the jammer can either (i) apply a simple threshold rule, activating transmission whenever observed metrics (e.g., bitrate or BLER) exceed a predefined level, or (ii) perform unsupervised traffic classification using DBSCAN to identify and react to specific traffic patterns. In both cases, the jammer selectively transmits during predicted active periods enabling dynamic, context-aware interference that reacts to evolving traffic patterns in the network. All traffic scenarios in our experiments in both Part A and Part B are generated using UDP-based \texttt{iperf3} traffic initiated from the base station to the UE, simulating downlink communication.

\begin{figure}[t]
    \centering
    \includegraphics[width=0.7\linewidth, height=1.7in]{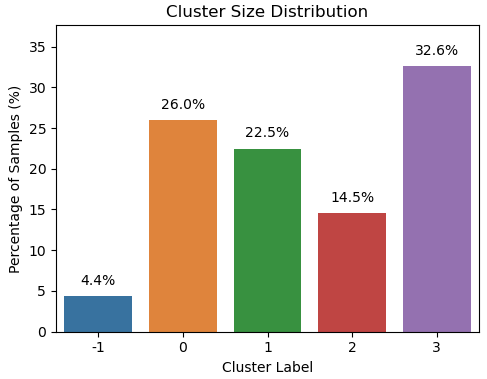}
    \caption{Distribution of samples across DBSCAN-generated clusters. Noise samples are denoted by label $-1$.}
\label{fig:cluster_dist}
\vspace{-0.22in}
\end{figure}

\subsection{Experimental Scenarios}\label{traf_sched}

To evaluate and compare jamming strategies under realistic and repeatable conditions, we design two controlled experimental scenarios, each representing distinct traffic behaviors and jamming constraints.

\paragraph{Part A: Fixed Traffic Scenario (Single Traffic Class, Unlimited Jamming Budget)}
This scenario models a UE with a consistent downlink demand, transmitting at a fixed bitrate of \textit{4 Mbps}. The traffic scheduler randomly alternates between active transmission periods and idle intervals. A fixed random seed is used to maintain deterministic traffic behavior across jammer evaluations. This setup enables fair comparison between jammers with no constraints on jamming duration. The total traffic session spans approximately \textbf{270 seconds}, with \textit{75.4\% active traffic} and \textit{24.6\% idle time}.

This scenario is used to compare three jamming strategies: a conventional \textit{Always-on jammer}, a \textit{Random jammer}, and a simple yet effective \textit{Y1-threshold jammer}. The goal is to demonstrate that under unlimited jamming budgets, a threshold-based jammer, enabled by Y1 analytics can match the interference impact of an always-on approach while significantly reducing transmission overhead by aligning jamming activity with actual traffic periods.

\paragraph{Part B: Multi-Rate Traffic Scenario with Budget-Constrained Jamming}
To evaluate jammer selectivity and intelligence under limited activity budgets, we create a diverse traffic profile that includes four traffic classes: \textit{4 Mbps (35.5\%)}, \textit{2 Mbps (24.5\%)}, \textit{500 Kbps (24.5\%)}, and \textit{Idle periods (15.5\%)}, across a total duration of \textbf{220 seconds}. This scenario is designed to demonstrate how a Y1-aided jammer can achieve greater disruption by selectively targeting high-impact traffic classes under constrained jamming budgets. 

The Y1-aided jammer first collects RAN analytics from the malicious Y1 consumer and applies unsupervised clustering (DBSCAN) offline to learn traffic patterns. During runtime, the jammer uses this traffic profile to activate jamming only when specific clusters are detected. We evaluate multiple jamming strategies derived from this cluster-aware model, including jammers that target high traffic, medium traffic, and low traffic clusters, and compare their effectiveness against a baseline random jammer across varying jamming budgets (10\%, 15\%, 20\%, 25\% of total session time). This setup allows us to investigate the disruption-efficiency tradeoff and highlight how targeting higher-throughput traffic yields significantly greater impact per unit transmission time.

% These scenarios enable a comprehensive study of baseline, random, threshold-based, and clustering-based jamming behaviors under both unconstrained and constrained operational settings.

% \begin{figure*}[t]
% \centering
% \begin{minipage}[t]{0.27\textwidth}
%     \centering
%     \includegraphics[width=\linewidth]{images/snr.png}
% \end{minipage}%
% \hfill
% \begin{minipage}[t]{0.27\textwidth}
%     \centering
%     \includegraphics[width=\linewidth]{images/brate.png}
% \end{minipage}%
% \hfill
% \begin{minipage}[t]{0.27\textwidth}
%     \centering
%     \includegraphics[width=\linewidth]{images/bler.png}
% \end{minipage}
% \vspace{-0.1in}
% \caption{Part A: Network Performance Metrics}
% \label{fig:network-performance}
% \vspace{-0.15in}
% \end{figure*}

\begin{figure*}[t]
\vspace{0.1in}
\centering
% --- Figure 4: Performance Metrics ---
\begin{minipage}[t]{0.72\textwidth}
    \centering
    \begin{minipage}[t]{0.3\textwidth}
        \centering
        \includegraphics[width=\linewidth]{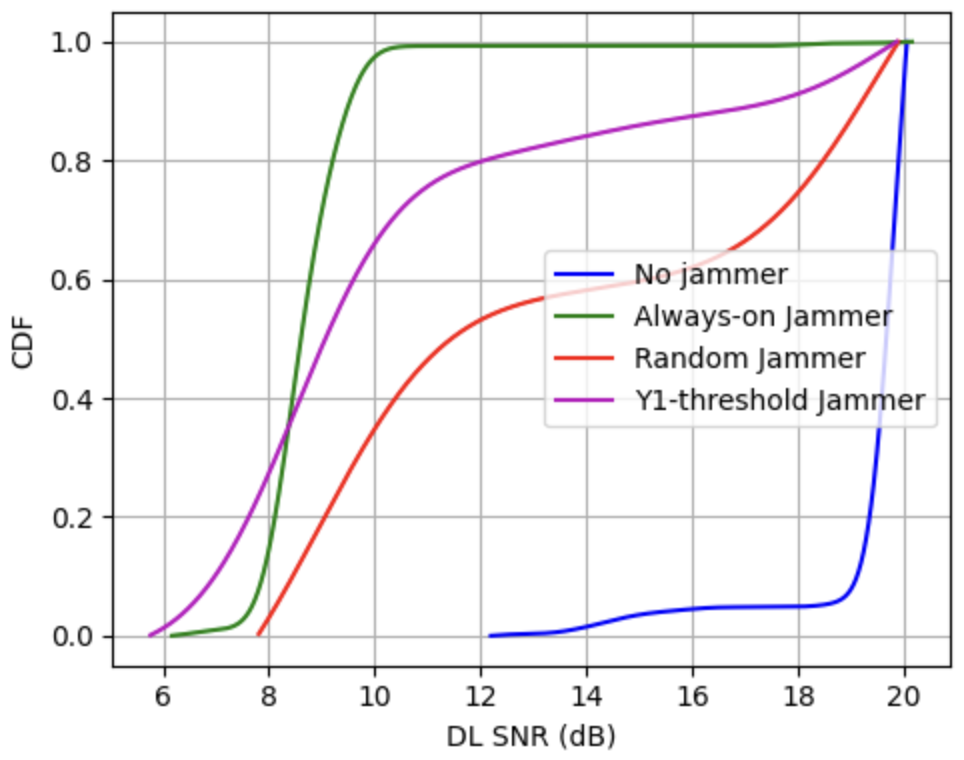}
    \end{minipage}%
    \hfill
    \begin{minipage}[t]{0.3\textwidth}
        \centering
        \includegraphics[width=\linewidth]{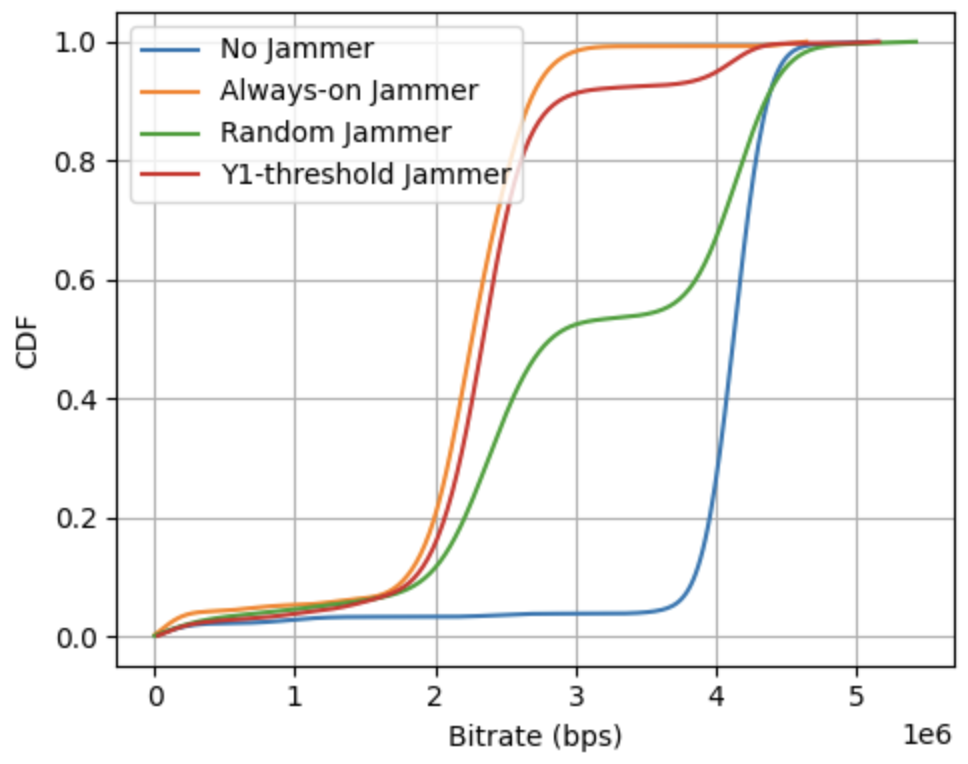}
    \end{minipage}%
    \hfill
    \begin{minipage}[t]{0.3\textwidth}
        \centering
        \includegraphics[width=\linewidth]{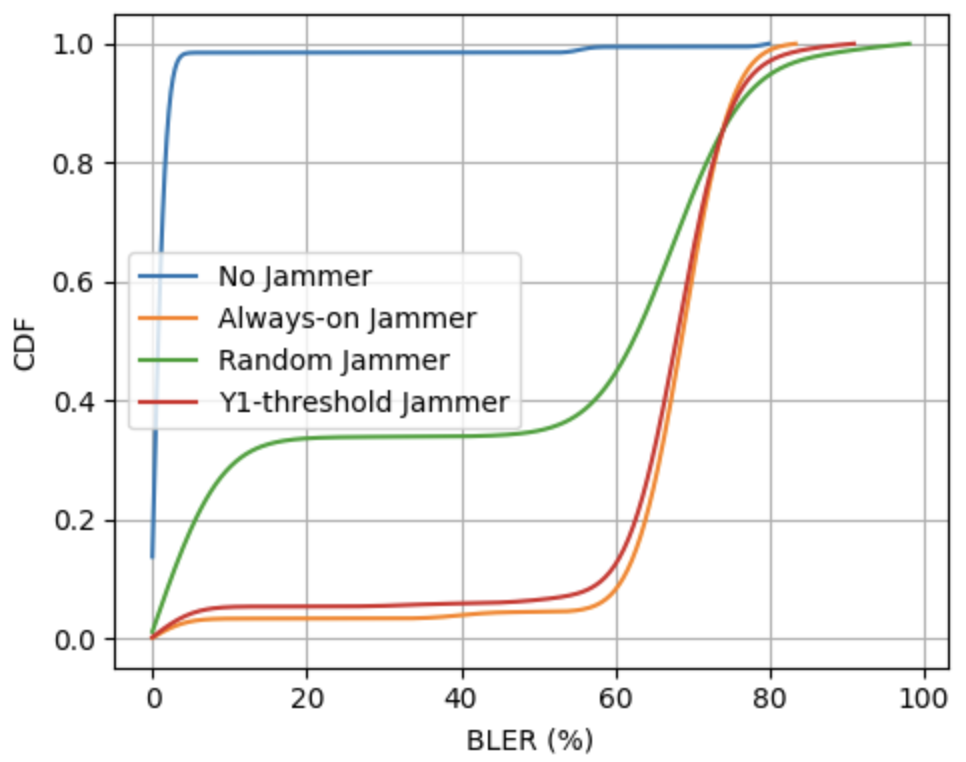}
    \end{minipage}
    \vspace{-0.1in}
    \caption{Part A: Network Performance Metrics}
    \label{fig:network-performance}
    \vspace{-0.3in}
\end{minipage}%
\hfill
% --- Figure 5: Bitrate Drop ---
\begin{minipage}[t]{0.25\textwidth}
    \centering
    \includegraphics[width=\linewidth,height=3.1cm]{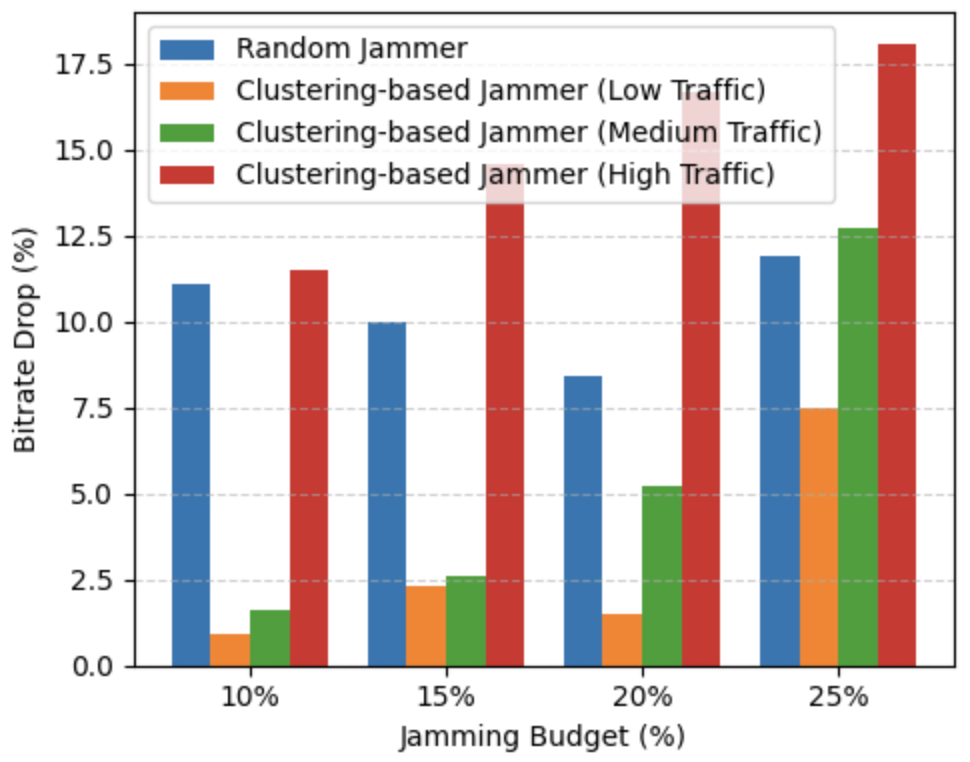}
    \vspace{-0.25in}
    \caption{Bitrate drop v.s. jamming budget}
    \label{fig:bitrate_drop_vs_budget}
\end{minipage}

\vspace{-0.25in}
\end{figure*}

\subsection{Clustering-Based Traffic Profiling for Adaptive Jamming}

To support live inference in the Part B experiments, we trained the DBSCAN model using $n = 227$ samples of RAN analytics, each characterized by a 4-dimensional feature vector: \texttt{[CQI, MCS, Bitrate, BLER]}. After parameter tuning, the optimal clustering was achieved with a neighborhood radius of $\varepsilon = 0.30$ and a minimum point count of $\texttt{minPts} = 10$. This configuration yielded four distinct clusters and a small noise group, as shown in Fig.~\ref{fig:cluster_dist}.

Manual inspection of the feature centroids revealed semantic groupings: Cluster 3 corresponded to high traffic (4 Mbps), Cluster 0 to medium traffic (500 Kbps), Cluster 2 to low traffic (100 Kbps), and Cluster 1 to idle states (near-zero activity). These labels were used to guide the jammer’s live decisions in Part B, enabling selective interference based on estimated traffic levels.

% \subsection{Evaluation Metrics}
\subsection{Network Performance Metrics}
To evaluate the impact of jamming on RAN performance, we focus on three key downlink performance indicators measured at the base station: \textbf{signal-to-noise ratio (SNR)}, \textbf{block error rate (BLER)}, and \textbf{throughput (DL bitrate)}. SNR reflects link quality, BLER quantifies transmission reliability, and throughput measures the effective data delivery rate under interference. Additionally, we introduce the \textbf{Bitrate Drop Percentage}, which quantifies the relative throughput degradation compared to the no-jamming baseline, it highlights the efficiency of a jammer in degrading the link quality. All metrics are averaged across active UEs (i.e., those with non-zero traffic) and visualized via cumulative distribution function (CDF) plots and tabular comparisons to evaluate interference effectiveness across strategies and jamming budgets.

\begin{table}[t]
\centering
\tiny
\caption{Part A Jammer Comparison (Single Traffic Class)}
\vspace{-0.1in}
\begin{tabular}{p{1.6cm}>{\centering}p{0.9cm}>{\centering}p{0.45cm}p{0.5cm}>{\centering}p{1cm}>{\centering\arraybackslash}p{1cm}}
\toprule
\textbf{Strategy} & \textbf{BLER (\%)} & \textbf{SNR (dB)} & \textbf{Bitrate (bps)} & \textbf{Bitrate Drop (\%)} & \textbf{Active Time (\%)} \\
\midrule
No Jammer         & 0.72 & 19.6 & 3949274.9 & 0 & 0     \\
Always-on Jammer  & 64.27 & 8.74 & 2154627.8 & 45.4 & 100   \\
Random Jammer     & 34.54 & 13.88 & 3025942.7 & 23.4 & 56   \\
Y1-Threshold Jammer  & 61.52 & 11.03 & 2348195.3 & 40.5 & 73 \\
\bottomrule
\end{tabular}
\label{tab:partA_jammer_comparison}
\vspace{-0.25in}
\end{table}

\subsection{Network Performance Results}
\subsubsection{Part A Result - Evaluation under Fixed Traffic and unlimited Jamming Budget} Table~\ref{tab:partA_jammer_comparison} summarizes the performance of different jamming strategies under a single-rate traffic profile. As expected, the \textit{Always-on jammer} introduces the most disruption, achieving the highest BLER (64.27\%) and the largest bitrate reduction (45.4\%) relative to the no-jamming case, but at the cost of continuous transmission (100\% active time). In contrast, the \textit{Y1-Threshold jammer}, guided by real-time traffic awareness, achieves nearly the same BLER (61.52\%) with only 73\% transmission activity closely aligned with the true traffic activity level of 75.4\% as defined in the traffic scenario. This confirms the effectiveness of using Y1-based jamming to selectively jam only during meaningful transmission windows, avoiding unnecessary interference during idle periods.

Although the SNR of the Y1-threshold jammer (11.03 dB) is slightly higher than that of the Always-on jammer (8.74 dB), this is due to its deactivation during idle periods when no jamming is required, allowing clear channel conditions. The \textit{Random jammer}, while active for 56\% of the time, delivers less disruption (BLER of 34.54\%, bitrate drop of 23.4\%), highlighting the benefit of targeted jamming over probabilistic activation. This behavior is also reflected in the CDF plots shown in Fig.~\ref{fig:network-performance}, where the Y1-threshold jammer closely mirrors the Always-on jammer in both BLER and bitrate metrics. An exception is observed in the SNR plot, where the threshold jammer achieves slightly higher SNR values due to its inactivity during idle periods, allowing for cleaner signal conditions compared to the Always-on jammer. In contrast, the Random jammer deviates more significantly across all metrics, highlighting its less effective and less targeted jamming behavior. These results demonstrate that even a simple Y1-threshold based jamming strategy can achieve near-optimal interference with significantly reduced energy and transmission effort, validating the threat potential of analytics-driven jammers.

\begin{table*}[ht]
\centering
\caption{Part B: Jamming Effectiveness Across Traffic Types and Budgets}
\vspace{-0.1in}
\label{tab:partB_results}
\resizebox{\textwidth}{!}{%
\begin{tabular}{lcccccccccccc}
\toprule
\textbf{Strategy} & \multicolumn{3}{c}{\textbf{10\% Budget}} & \multicolumn{3}{c}{\textbf{15\% Budget}} & \multicolumn{3}{c}{\textbf{20\% Budget}} & \multicolumn{3}{c}{\textbf{25\% Budget}} \\
\cmidrule(lr){2-4} \cmidrule(lr){5-7} \cmidrule(lr){8-10} \cmidrule(lr){11-13}
 & Bitrate & Drop (\%) & BLER & Bitrate & Drop (\%) & BLER & Bitrate & Drop (\%) & BLER & Bitrate & Drop (\%) & BLER \\
\midrule
No Jammer     & 2418944 & --     & 0.06  & 2418944 & --     & 0.06  & 2418944 & --     & 0.06  & 2386517 & --     & 0.03 \\
Random Jammer       & 2151102 & 11.1   & 5.86  & 2177414 & 10.0   & 12.34 & 2215156 & 8.4    & 15.09 & 2102078 & 11.9   & 14.61 \\
Clustering-Based Jammer (Low Traffic)   & 2398083 & 0.9    & 7.90  & 2363917 & 2.3    & 12.04 & 2383796 & 1.5    & 13.98 & 2208256 & 7.5    & 21.78 \\
Clustering-Based Jammer (Medium Traffic) & 2380269 & 1.6    & 6.47  & 2355697 & 2.6    & 12.78 & 2294049 & 5.2    & 17.17 & 2082571 & 12.7   & 20.80 \\
\textbf{Clustering-Based Jammer (High Traffic)}  & \textbf{2141635} & \textbf{11.5} & 8.33  & \textbf{2065406} & \textbf{14.6} & 10.61 & \textbf{2013937} & \textbf{16.7} & 20.71 & \textbf{1954988} & \textbf{18.1} & 18.47 \\
\bottomrule
\end{tabular}%
}
\vspace{-0.2in}
\end{table*}

% \begin{figure}[t]
%   \vspace{-0.1in}
%   \centering
%   \includegraphics[width=0.32\textwidth, height=4.2cm]{images/brate_drop.png}
%     \vspace{-0.15in}
%   \caption{Bitrate Drop Across the Jamming Budget}
%   \label{fig:bitrate_drop_vs_budget}
%   \vspace{-0.28in}
% \end{figure}

\subsubsection{Part B: Adaptive Jamming with Budget Constraints}
Fig.\ref{fig:bitrate_drop_vs_budget} and Table~\ref{tab:partB_results} evaluate the disruption potential of various Y1-aided jamming strategies under constrained jamming budgets (10\% to 25\% of total session time). This experiment was designed to demonstrate how an intelligent jammer, equipped with prior traffic profiling capabilities, can maximize disruption while consuming the same jamming budget as a less strategic adversary. The random jammer, which lacks context-awareness, consistently achieves moderate bitrate degradation across all budgets. Despite its unpredictable activation pattern, it often outperforms jammers that target lower-rate traffic (i.e., 500 Kbps or 2 Mbps), which tend to carry less data and thus contribute less to overall throughput reduction when disrupted compared to targeting higher-rate traffic.

In contrast, as shown in Fig.\ref{fig:bitrate_drop_vs_budget}, the High Traffic jammer, guided by unsupervised traffic clustering, consistently delivers the highest disruption efficiency. It achieves a maximum bitrate drop of \textbf{18.1\%} at 25\% jamming budget, substantially higher than the random and lower traffic jammers, by focusing its transmission solely on the highest-bandwidth flow (4 Mbps). This confirms that targeting high-throughput periods maximizes disruption per unit of transmission time, validating the benefits of profiling-aware jamming.

Interestingly, the BLER values for High Traffic jamming are not always the highest. This counterintuitive trend likely results from rapid link adaptation in response to aggressive jamming during high MCS usage, which forces the network to lower its modulation and coding scheme to preserve reliability. As a result, throughput suffers significantly while average BLER remains controlled.

Overall, these results highlight the advantage of Y1-guided traffic profiling in enabling strategic, cost-effective jamming. When transmission budget is limited as would be the case in energy-constrained or covert scenarios intelligent selection of jamming windows can amplify disruption impact without requiring full-time interference.

\subsection{Discussion}
In this work, we implemented and demonstrated a new attack surface that emerges from the exposure of RAN analytics via the O-RAN Y1 interface. While our experimental framework employed a set of carefully selected RAN metrics shown in Table \ref{tab:analytics-metrics} that are not explicitly listed in the current Y1 specification, we emphasize that our choice aligns with the broader goal of simulating meaningful and realistic telemetry exchange between the RAN and authorized consumers. These metrics reflect aggregate downlink behavior, avoiding per-UE granularity, and were selected based on their prevalence in typical baseband telemetry. Moreover, these metrics are representative of the types of network characteristics captured by analytics listed in the YI specification, such as average packet loss, throughput, and packet delay rate. Importantly, our methodology and attack model are agnostic to the exact analytic fields and could be extended to support the analytics explicitly defined in the Y1 specification without altering the overall flow or inference mechanism.

Our findings underscore a key insight: the presence of any sufficiently informative RAN telemetry can serve as a potent enabler for adversarial profiling. Even limited exposure to periodic or aggregated statistics allows a well-positioned attacker to learn temporal traffic patterns, and selectively trigger interference with minimal footprint. 

\section{Conclusion and Future Work}
% In this work, we evaluated the feasibility and effectiveness of jamming attacks guided by RAN analytics shared over the O-RAN Y1 interface. We developed and analyzed two Y1-aided jammer strategies -- Y1-threshold jammer, and DBSCAN clustering-based jammer against two baselines, namely always-on and random jammers, using an OTA O-RAN testbed. Our experiments show that even simple Y1-threshold jammers can mirror the effectiveness of full-time, always-on jamming while reducing transmission overhead by over 25\%. When constrained by jamming budgets, profiling-driven strategies, i.e., clustering-based jammers, that target high-bandwidth traffic significantly outperform both random and poorly targeted attacks achieving up to 18\% throughput degradation using only 25\% transmission time.

% These results underscore the potential for malicious actors to exploit open interfaces like Y1 for stealthy, efficient and smarter jamming, raising security concerns for future O-RAN based 5G/6G networks, especially in challenging military  environments. Future work will explore defenses such as access control at the Y1 interface.

This work demonstrates that jamming attacks guided by RAN analytics from the O-RAN Y1 interface can be highly effective and efficient. Using an OTA testbed, two Y1-aided strategies, a threshold-based jammer and a DBSCAN-based jammer, were evaluated against always-on and random jammers. Results show that even lightweight Y1-based jammers can match the disruption of always-on attacks while cutting transmission overhead by over 25\%. Clustering-based jammers further improve efficiency, achieving up to 18\% throughput degradation with only 25\% transmission time. These highlight the risks of exposing analytics via open interfaces like Y1, especially in military contexts, motivating future research on Y1 access control defenses. 

\section{Acknowledgment}
This work has been supported by the Public Wireless Supply Chain Innovation Fund (PWSCIF) under Federal Award ID Number 51-60-IF007, NSF award 2120411, 2145631, and the NSF and Office of the Under Secretary of Defense Research and Engineering ITE 2326898 and 2515378, as part of the NSF Convergence Accelerator Track G: Securely Operating Through 5G Infrastructure Program.

\bibliographystyle{IEEEtran}
\bibliography{ref}

\end{document}